\documentclass[12pt]{article}

\usepackage{amsmath}
\usepackage{cite}
\usepackage[xy,curve,matrix,oldcm,debug]{qsymbols}
\usepackage{hhline}

\def\nn{\nonumber}

\numberwithin{equation}{section}

\title{Lagrangian description of the partially massless\\
 higher spin $N=1$ supermultiplets in $AdS_4$ space}

\author{I.L. Buchbinder${}^{ab}$\thanks{joseph@tspu.edu.ru},
M.V. Khabarov${}^{cd}$\thanks{maksim.khabarov@ihep.ru}, T.V.
Snegirev${}^{ae}$\thanks{snegirev@tspu.edu.ru}, Yu.M.
Zinoviev${}^{cd}$\thanks{Yurii.Zinoviev@ihep.ru}
\\[0.5cm]
\it{\small ${}^a$Department of Theoretical Physics, Tomsk State
Pedagogical University,}\\
\it{\small Tomsk, 634061, Russia}\\
\it{\small ${}^b$National Research Tomsk State University, Tomsk
634050, Russia}\\
\it{\small ${}^c$Institute for High Energy Physics of National
Research Center "Kurchatov Institute"} \\
\it{\small Protvino, Moscow Region, 142281, Russia} \\
\it\small{ ${}^d$Moscow Institute of Physics and Technology (State
University),} \\
\it{\small Dolgoprudny, Moscow Region, 141701, Russia}\\
\it{\small ${}^e$National Research Tomsk Polytechnic University,
Tomsk 634050, Russia}}

\date{}

\begin{document}

\maketitle

\begin{abstract}
In the recent paper [1] the classification of non-unitary
representations of the three dimensional superconformal group has
been constructed. From $AdS/CFT$ they must correspond to
$N=1$ supermultiplets containing partially massless fields in
$AdS_4$. Moreover, the simplest example of such supermultiplets
which contains a partially massless spin-$2$ was explicitly
constructed. In this paper we extend this result and develop
explicit Lagrangian construction of general $N=1$ supermultiplets
containing partially massless fields with arbitrary superspin. We
use the frame-like gauge invariant description of partially massless
higher spin bosonic and fermionic fields. For the two types of the
supermultiplets (with integer and half-integer superspins) each one
containing two partially massless bosonic and two partially massless
fermionic fields we derive the supertransformations leaving the sum
of four their free Lagrangians invariant such that the $AdS_4$
superalgebra is closed on-shell.
\end{abstract}

\thispagestyle{empty}
\newpage
\setcounter{page}{1}
\tableofcontents\pagebreak

\section{Introduction}

In the recent paper \cite{GHR18} the classification of non-unitary
representations of the three dimensional superconformal group has
been constructed. From $AdS/CFT$-correspondence (i.e. from the fact
that the very same superalgebra plays a  role of the super-$AdS_4$
algebra in the bulk and of the superconformal one on the boundary)
there must exist their analogues in four-dimensional Anti de Sitter
space ($AdS_4$) as well. By the structure of the supermultiplets they
constructed, the authors of \cite{GHR18} suggested that they
correspond to the supermultiplets with the partially massless fields
which are also non-unitary in $AdS_4$. Moreover, the
simplest example of such supermultiplets which contains a partially
massless spin-$2$, massless spin-$1$, massless spin-$3/2$ and
massive spin-$3/2$ was explicitly constructed. The dynamical
description of the arbitrary supermultiplets was not studied. In
this paper we fill this gap and construct explicit Lagrangian
realization of all $N=1$ supermultiplets containing partially
massless fields with arbitrary integer and half-integer superspins.

The partially massless fields \cite{DW01,DW01a,Zin01,Met06,SV06}
(non-unitary in $AdS$) of integer $s$ or half-integer $s+1/2$ spins
are labelled by depth $t\in\{0,1,2,...,(s-1),s\}$. Two boundary
values $t=0$ and $t=s$ correspond to massless and massive cases
respectively. For other values of $t$ we have pure partially
massless field which propagates $2(t+1)$ degrees of freedom. As it
was shown in \cite{GHR18} the general partially massless $N=1$
supermultiplets are described by the diagrams
\begin{equation}\label{PMSM}
\xymatrix {& {[s+\frac12]_{t}} \ar@{-}[dl]_-{} \ar@{-}[dr]^-{} &\\
 {[s]_{t}} \ar@{-}[dr]_-{} & Y=s & {[{s}]_{t-1}}\\
 & {[s-\frac12]_{t-1}} \ar@{-}[ur]_-{} & }\qquad
 \xymatrix {& {[s-\frac12]_{t}} \ar@{-}[dl]_-{} \ar@{-}[dr]^-{} &\\
 {[s]_{t}} \ar@{-}[dr]_-{} & Y=s-\frac12 & {[{s-1}]_{t-1}}\\
 & {[s-\frac12]_{t-1}} \ar@{-}[ur]_-{} & }
\end{equation}
Here integers $s$ and $t$ are the spin and the depth of partially
massless fields. As in the massive case $N=1$ partially-massless
supermultiplets contain a pair of the bosonic fields and a pair of the
fermionic ones. For instance, left diagram describes partially
massless supermultiplet with superspin $Y=s$ which contains two
partially massless bosonic spin-$s$ fields of depth $t$ and $(t-1)$,
partially massless fermionic spin-$(s+1/2)$ field of depth $t$ and
partially massless fermionic spin-$(s-1/2)$ field of depth $(t-1)$.
Taking into account that depth $t$ partially massless fields propagate
$2(t+1)$ degrees of freedom it is easy to check that the number of
bosonic and fermionic degrees of freedom matches and equals $4t$.

For the description of the individual partially massless higher spin
bosonic and fermionic fields we use the frame-like gauge invariant
description similar to the massive case \cite{Zin08b,PV10,BKSZ18a}.
In such formalism partially massless spin $s(s+1/2)$ of depth $t$ is
described by a set of massless fields with spins $s,s-1,...,s-t$
combined together into one system\footnote{As in the massive case the
fields with spins $s-1,...,s-t$ are auxiliary and play a role of the
Stuckelberg fields. In works \cite{BS02,HW05} it was shown that in the
metric-like formalism they can be derived from a log radial
dimensional reduction of massless theory. In the case of the
frame-like formalism we used, in general the reduction produces more
field components than it is necessary so one has to exclude the
unnecessary ones by solving their equations and/or gauge fixing. This
is even more true for the supermultiplets because starting with
$N=1$ supersymmetry in higher dimensions one usually ends with the
$N=2$ supersymmetry and again has to truncate somehow to go back to
$N=1$.}. To combine partially massless fields into supermultiplets
(\ref{PMSM}) we follow the strategy of our recent paper \cite{BKSZ18a}
where massive higher spin supermultiplets were constructed. For the
Lagrangian we just take the sum of four free Lagrangians for the two
partially massless bosonic and two partially massless fermionic fields
entering the supermultiplet. Then for each pair of bosonic and
fermionic fields (we call it superblock in what follows) we find the
supertransformations leaving the sum of their two Lagrangians
invariant. After that we combine all four possible superblocks and
adjust their parameters so that the algebra of the
supertransformations be closed on-shell.

The paper is organized as follows. In section \ref{Section2} we give
non-unitary frame-like gauge invariant formulation for free
partially massless arbitrary integer and half-inter spins. In
section \ref{Section3} we consider superblocks containing one
partially massless bosonic and one partially massless fermionic
fields and find corresponding supertransformations. In section
\ref{Section4} we combine the constructed partially massless
superblocks into the partially massless supermultiplets.

\section{Partially massless higher spin fields}\label{Section2}

In this section we provide frame-like gauge invariant formulation for
(non-unitary) partially massless fields with arbitrary integer and
half-inter spins in $AdS_4$ space.

\subsection{Partially massless bosons}\label{PMBoson}

The gauge invariant formulation for the partially massless fields
can be easily obtained from the general massive case just by
adjusting the value of mass parameter. But unitarity requires that
the sign of the cosmological term be positive so that naturally the
partially massless fields live in de Sitter space. In this work we
use the gauge invariant formulation for the partially massless
fields in $AdS_4$ space where half the number of components have
wrong signs of the kinetic terms. Such a description is explicitly
non-unitary but the Lagrangian is hermitian and all coefficients are
real.

In such approach a partially massless integer spin-$s$ field of
depth $t=(s-l-1)$ is formulated in terms of massless fields with
spins $(l+1)\leq k\leq s$. Each massless bosonic fields with spin
$k\geq2$ (the case of the partially massless bosonic fields of the
last depth $t=(s-1)$ requires introduction of the spin-1 component
and has to be considered separately) described by the physical
one-form $f^{\alpha(k-1)\dot\alpha(k-1)}$ and the auxiliary one-forms
$\Omega^{\alpha(k)\dot\alpha(k-2)},\Omega^{\alpha(k-2)\dot\alpha(k)}$.
They are two-component multispinors symmetric on its local dotted
and undotted spinorial indices separately. These fields satisfy the
following reality condition
\begin{equation}\label{RealCond}
(f^{\alpha(k-1)\dot\alpha(k-1)})^\dag =
f^{\alpha(k-1)\dot\alpha(k-1)},\qquad
(\Omega^{\alpha(k)\dot\alpha(k-2)})^\dag =
\Omega^{\alpha(k-2)\dot\alpha(k)}.
\end{equation}

In these notations the gauge invariant Lagrangian for the partially
massless bosonic field can be written as follows:
\begin{eqnarray}\label{PMBosonLag}
(-1)^\sigma\frac{1}{i}{\cal L} &=&
\sum_{k=l+1}^{s}[k\Omega^{\alpha(k-1)\beta\dot\alpha(k-2)}
E_\beta{}^\gamma
\Omega_{\alpha(k-1)\gamma\dot\alpha(k-2)} \nonumber
\\
&& \qquad -(k-2)\Omega^{\alpha(k)\dot\alpha(k-3)\dot\beta}
E_{\dot\beta}{}^{\dot\gamma}
\Omega_{\alpha(k)\dot\alpha(k-3)\dot\gamma} \nonumber
\\
&& \qquad +2\Omega^{\alpha(k-1)\beta\dot\alpha(k-2)}
e_\beta{}^{\dot\beta}
Df_{\alpha(k-1)\dot\alpha(k-2)\dot\beta}-h.c.]\nonumber
\\
&& +\sum_{k=l+2}^{s}a_k[E_{\beta(2)}
\Omega^{\alpha(k-2)\beta(2)\dot\alpha(k-2)}
f_{\alpha(k-2)\dot\alpha(k-2)} \nonumber
\\
&& \qquad \qquad + \frac{(k-2)}{k}E_{\beta(2)}
f^{\alpha(k-3)\beta(2)\dot\alpha(k-1)}
\Omega_{\alpha(k-3)\dot\alpha(k-1)}-h.c.] \nonumber
\\
&& +\sum_{k=l+1}^sb_k [f^{\alpha(k-2)\beta\dot\alpha(k-1)}
E_\beta{}^\gamma f_{\alpha(k-2)\gamma\dot\alpha(k-1)}-h.c.].
\end{eqnarray}
Here the even/odd parameter $\sigma$ determines the common sign of
the Lagrangian that will be important for the construction of the
supermultiplets. The Lagrangian (\ref{PMBosonLag}) is invariant
under the following gauge transformations:
\begin{eqnarray}\label{PMBosonGT}
\delta
f^{\alpha(k-1)\dot\alpha(k-1)} &=& D\xi^{\alpha(k-1)\dot\alpha(k-1)}
+e_\beta{}^{\dot\alpha} \eta^{\alpha(k-1)\beta\dot\alpha(k-2)} +
e^\alpha{}_{\dot\beta} \eta^{\alpha(k-2)\dot\alpha(k-1)\dot\beta}
\nonumber
\\
&& -\frac{(k-1)a_{k+1}}{2(k+1)} e_{\beta\dot\beta}
\xi^{\alpha(k-1)\beta\dot\alpha(k-1)\dot\beta}
+ \frac{a_k}{2k(k-1)} e^{\alpha\dot\alpha}
\xi^{\alpha(k-2)\dot\alpha(k-2)} \nonumber
\\
\delta \Omega^{\alpha(k)\dot\alpha(k-2)} &=&
D\eta^{\alpha(k),\dot\alpha(k-2)} -\frac{a_{k+1}}{2}
e_{\beta\dot\beta} \eta^{\alpha(k)\beta\dot\alpha(k-2)\dot\beta}
\\
&& +\frac{a_k}{2k(k+1)} e^{\alpha\dot\alpha}
\eta^{\alpha(k-1)\dot\alpha(k-3)} + \frac{b_k}{2k}
e^\alpha{}_{\dot\beta} \xi^{\alpha(k-1)\dot\alpha(k-2)\dot\beta},
\nonumber
\end{eqnarray}
provided
\begin{eqnarray}\label{boson_date}
b_k &=& \frac{2s(s+1)l(l+1)}{k(k-1)(k+1)}\lambda^2 \nonumber
\\
a_k{}^2 &=& \frac{4(s-k+1)(s+k)(k-l-1)(k+l)}{(k-2)(k-1)}\lambda^2.
\end{eqnarray}
In what follows we assume that all parameters $a_k$ are positive. It
is also worth to note that Lagrangian (\ref{PMBosonLag}) is parity
invariant that is invariant under spatial reflections. These
transformations can be defined by operator $P$ as follows
\begin{eqnarray}\label{Peven}
Pf^{\alpha(k-1)\dot\alpha(k-1)}=
f^{\alpha(k-1)\dot\alpha(k-1)},\quad
P\Omega^{\alpha(k)\dot\alpha(k-2)}=
\Omega^{\alpha(k-2)\dot\alpha(k)},
\end{eqnarray}
$$
Pe^{\alpha\dot\alpha}=e^{\alpha\dot\alpha},\quad
PE^{\alpha\beta}=E^{\dot\alpha\dot\beta}.
$$
Using the fact that Lagrangian in four dimensions is differential
4-form which is proportional to antisymmetric tensor
$\varepsilon_{\mu\nu\rho\sigma}$ and
$P\varepsilon_{\mu\nu\rho\sigma}=-\varepsilon_{\mu\nu\rho\sigma}$ we
can see that Lagrangian (\ref{PMBosonLag}) is $P$-invariant.
Moreover, due to the Lagrangian is quadratic in fields, it describes
both parity-even boson defined by (\ref{Peven}) and parity-odd one
defined by
\begin{eqnarray}\label{Podd}
Pf^{\alpha(k-1)\dot\alpha(k-1)}=-
f^{\alpha(k-1)\dot\alpha(k-1)},\quad
P\Omega^{\alpha(k)\dot\alpha(k-2)}=-
\Omega^{\alpha(k-2)\dot\alpha(k)}.
\end{eqnarray}

In the gauge invariant formalism we use, for each field (physical or
auxiliary) there exist a corresponding gauge invariant object
("torsion" or "curvature"). Their form is completely determined by the
structure of the gauge transformations (\ref{PMBosonGT})\footnote{Note
that to construct a full set of gauge invariant objects one has to
introduce a number of so-called extra fields. But these fields do no
enter the free Lagrangian so in what follows we omit them.}:
\begin{eqnarray}\label{PMBosonCurv}
{\cal T}^{\alpha(k-1)\dot\alpha(k-1)} &=&
Df^{\alpha(k-1)\dot\alpha(k-1)} +
e_\beta{}^{\dot\alpha}\Omega^{\alpha(k-1)\beta\dot\alpha(k-2)} +
e^\alpha{}_{\dot\beta}\Omega^{\alpha(k-2)\dot\alpha(k-1)\dot\beta}
\nonumber
\\
&& - \frac{(k-1)a_{k+1}}{2(k+1)} e_{\beta\dot\beta}
f^{\alpha(k-1)\beta\dot\alpha(k-1)\dot\beta} +
\frac{a_k}{2k(k-1)}e^{\alpha\dot\alpha} 
f^{\alpha(k-2)\dot\alpha(k-2)}, \nonumber
\\
{\cal R}^{\alpha(k)\dot\alpha(k-2)} &=&
D\Omega^{\alpha(k),\dot\alpha(k-2)} - \frac{a_{k+1}}{2}
e_{\beta\dot\beta} \Omega^{\alpha(k)\beta\dot\alpha(k-2)\dot\beta}
\\
&& + \frac{a_k}{2k(k+1)} e^{\alpha\dot\alpha}
\Omega^{\alpha(k-1)\dot\alpha(k-3)} + \frac{b_k}{2k}
e^\alpha{}_{\dot\beta} f^{\alpha(k-1)\dot\alpha(k-2)\dot\beta}.
\nonumber
\end{eqnarray}
In this work we use a formalism analogous to the so-called 1 and 1/2
order formalism, very well known in supergravity. Namely, we do not
introduce any supertransformations for the auxiliary fields, instead
all calculations are done using the "zero torsion conditions":
\begin{eqnarray}\label{PMOn-Shell}
{\cal T}^{\alpha(k-1)\dot\alpha(k-1)} \approx 0 &\Rightarrow&
e_\beta{}^{\dot\alpha} {\cal R}^{\alpha(k-1)\beta\dot\alpha(k-2)} +
e^\alpha{}_{\dot\beta} {\cal
R}^{\alpha(k-2)\dot\alpha(k-1)\dot\beta}\approx 0.
\end{eqnarray}
As for the supertransformations for the physical fields, the variation
of the Lagrangian can be compactly written using the gauge invariant
curvatures given above:
\begin{eqnarray}\label{BosonVar}
\delta{\cal L}&=&-(-1)^\sigma 2i\sum_{k=l+1}^s {\cal
R}^{\alpha(k-1)\beta\dot\alpha(k-2)} e_\beta{}^{\dot\beta} \delta
f_{\alpha(k-1)\dot\alpha(k-2)\dot\beta} - h.c.
\end{eqnarray}

\subsection{Partially massless fermions}\label{PMFerm1}

To construct a gauge invariant Lagrangian for the fermionic fields one
only needs physical fields. So to describe partially massless spin
s+1/2 field of depth $t = (s-l-1)$ we introduce a set of one-forms
$\Phi^{\alpha(k)\dot\alpha(k-1)},\Phi^{\alpha(k-1)\dot\alpha(k)}$,
$l+1 \le k \le s$ which are symmetric on their dotted and
undotted spinorial indices separately and satisfying a reality
condition
$$
(\Phi^{\alpha(k)\dot\alpha(k-1)})^\dag =
\Phi^{\alpha(k-1)\dot\alpha(k)}.
$$
The Lagrangian for the partially massless fields in $AdS_4$ has the
form
\begin{eqnarray}\label{FermLag}
(-1)^\tau{\cal L} &=& \sum_{k=l+1}^{s}
\Phi_{\alpha(k-1)\beta\dot\alpha(k-1)} e^\beta{}_{\dot\beta}
D\Phi^{\alpha(k-1)\dot\alpha(k-1)\dot\beta} \nonumber
\\
&& + \sum_{k=l+2}^{s}c_k[E^{\beta(2)}
\Phi_{\alpha(k-2)\beta(2)\dot\alpha(k-1)}
\Phi^{\alpha(k-2)\dot\alpha(k-1)} + h.c.] \nonumber
\\
&& + \sum_{k=l+1}^sd_{k}[(k+1)
\Phi_{\alpha(k-1)\beta\dot\alpha(k-1)}E^\beta{}_{\gamma}
\Phi^{\alpha(k-1)\gamma\dot\alpha(k-1)} \nonumber
\\
&& \qquad \qquad - (k-1)\Phi_{\alpha(k)\dot\alpha(k-2)\dot\beta}
E^{\dot\beta}{}_{\dot\gamma}
\Phi^{\alpha(k)\dot\alpha(k-2)\dot\gamma}+h.c.].
\end{eqnarray}

As in the bosonic case half the number of components have wrong signs
of the kinetic terms. Such a description is explicitly non-unitary but
the Lagrangian is hermitian and all coefficients are real. In what
follows we assume that the parameters $c_k$ are positive while
$\tau$ (even/odd) in Lagrangian (\ref{FermLag}) parameterize the
common sign of the Lagrangian.

This Lagrangian is invariant under the following gauge transformation:
\begin{eqnarray}
\delta\Phi^{\alpha(k)\dot\alpha(k-1)} &=&
D\xi^{\alpha(k)\dot\alpha(k-1)} + e_\beta{}^{\dot\alpha}
\eta^{\alpha(k)\beta\dot\alpha(k-2)} + 2d_{k}
e^\alpha{}_{\dot\beta} \xi^{\alpha(k-1)\dot\alpha(k-1)\dot\beta}
\nonumber
\\
&& - c_{k+1}e_{\beta\dot\beta}
\xi^{\alpha(k)\beta\dot\alpha(k-1)\dot\beta}
+ \frac{c_k}{(k-1)(k+1)} e^{\alpha\dot\alpha}
\xi^{\alpha(k-1)\dot\alpha(k-2)},
\end{eqnarray}
provided
\begin{eqnarray}\label{fermion_data}
d_k &=& \pm\frac{(s+1)(l+1)}{2k(k+1)}\lambda \nonumber
\\
c_k{}^2 &=& \frac{(s-k+1)(s+k+1)(k-l-1)(k+l+1)}{k^2}\lambda^2.
\end{eqnarray}
We assume that all parameters $c_k$ are positive. The sign of $d_k$
(which is not fixed by the gauge invariance) plays an important role
in the construction of the supermultiplets. As it will be seen below
the pair of the fermions entering $N=1$ supermultiplet must have
opposite signs of $d_k$ forming in this way the Dirac mass-like term.

In the fermionic case for each field we also have a corresponding
gauge invariant object (as in the bosonic case we omit any extra
fields):
\begin{eqnarray}\label{FermCurv}
{\cal F}^{\alpha(k)\dot\alpha(k-1)} &=&
D\Phi^{\alpha(k)\dot\alpha(k-1)} + 2d_{k}
e^\alpha{}_{\dot\beta}\Phi^{\alpha(k-1)\dot\alpha(k-1)\dot\beta}
\nonumber
\\
&& - c_{k+1}e_{\beta\dot\beta}
\Phi^{\alpha(k)\beta\dot\alpha(k-1)\dot\beta} +
\frac{c_k}{(k-1)(k+1)}e^{\alpha\dot\alpha}
\Phi^{\alpha(k-1)\dot\alpha(k-2)}.
\end{eqnarray}
Using these curvatures, the variation of the Lagrangian
(\ref{FermLag}) under the supertransformations can be compactly
written as follows:
\begin{equation}\label{FermVar}
\delta{\cal L} = - (-1)^\tau \sum_{k=\tilde{l}+1}^s {\cal
F}_{\alpha(k-1)\beta\dot\alpha(k-1)} e^\beta{}_{\dot\beta}
\delta\Phi^{\alpha(k-1)\dot\alpha(k-1)\dot\beta} + h.c.
\end{equation}

\section{Partially massless superblocks}\label{Section3}

As it has been shown in \cite{GHR18}, the partially massless
supermultiplets in $AdS_4$, corresponding to non-unitary
supersymmetric representations, similarly to the massive case
contain two bosonic and two fermionic partially massless fields with
the properly adjusted depths (see diagrams (\ref{PMSM}) in
Introduction):
$$
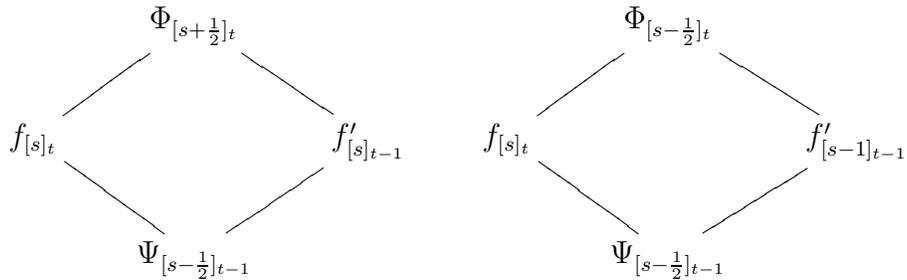
\xymatrix{  & \Phi_{[s+\frac12]_t} \ar@{-}[dr] &  \\
f_{[s]_t} \ar@{-}[ur] & & f'_{[s]_{t-1}} \ar@{-}[dl] \\
 & \Psi_{[s-\frac12]_{t-1}} \ar@{-}[ul] } \qquad
\xymatrix{  & \Phi_{[s-\frac12]_t} \ar@{-}[dr] &  \\
f_{[s]_t} \ar@{-}[ur] & & f'_{[s-1]_{t-1}} \ar@{-}[dl] \\
 & \Psi_{[s-\frac12]_{t-1}} \ar@{-}[ul] }
$$
Here integers $s$ and $t$ label spin and depth respectively. For
given $s$ the depth $t$ of bosonic (fermionic) partially massless
field with spin $s(s+1/2)$ go from 1 to $(s-1)$. The authors of
\cite{GHR18} also have fined Lagrangian realization of the simplest
supermultiplets containing partially massless spin-2, massive
spin-3/2 and two massless fields with spin 3/2 and spin 1 (it arises
from the right diagram at $s=2$ and $t=1$). They have studied it
from the partially massless limit of the full massive
supermultiplet. Such limit in AdS is non-unitary and lead to that
norms of kinetic terms of spin-2 and spin-1 fields in Lagrangian are
opposite, the same holds for two spin-3/2 fields.

In this work we systematically study generic partially massless
supermultiplets corresponding to the above diagrams for $s>2$ and
$1\leq t<(s-1)$ working from the beginning with the partially massless
fields. We follow the same strategy we used for the construction of
the massive supermultiplets \cite{BKSZ18a}. At first, we consider two
possible pairs of the bosonic and fermionic partially massless fields
(superblocks), namely $(s,s+1/2)$ and $(s-1/2,s)$, and find the
supertransformations which leave the sum of their free Lagrangians
invariant. Then we consider the whole system of four fields and
choose the parameters in such a way that the algebra of the
supertransformations is closed.

\subsection{Ansatz for the supertransformations}

We choose the following ansatz for the supertransformations for a
pair of the partially massless bosonic and fermionic fields
(superblock):
\begin{eqnarray}\label{PMST}
\delta f^{\alpha(k-1)\dot\alpha(k-1)} &=&
\alpha_{k-1}\Phi^{\alpha(k-1)\beta\dot\alpha(k-1)} \zeta_\beta -
\bar\alpha_{k-1}\Phi^{\alpha(k-1)\dot\alpha(k-1)\dot\beta}
\zeta_{\dot\beta} \nonumber\\
 && + \alpha'_{k-1} \Phi^{\alpha(k-1)\dot\alpha(k-2)}
\zeta^{\dot\alpha} - \bar\alpha'_{k-1}
\Phi^{\alpha(k-2)\dot\alpha(k-1)} \zeta^{\alpha}, \nonumber
\\
\delta \Phi^{\alpha(k)\dot\alpha(k-1)} &=&
\beta_{k-1}\Omega^{\alpha(k)\dot\alpha(k-2)} \zeta^{\dot\alpha} +
\gamma_{k-1}f^{\alpha(k-1)\dot\alpha(k-1)} \zeta^{\alpha} \nonumber
\\
 && + \beta'_{k} \Omega^{\alpha(k)\beta\dot\alpha(k-1)}
\zeta_\beta + \gamma'_{k}
f^{\alpha(k)\dot\alpha(k-1)\dot\beta}\zeta_{\dot\beta},
\\
\delta \Phi^{\alpha(k-1)\dot\alpha(k)} &=&
\bar\beta_{k-1}\Omega^{\alpha(k-2)\dot\alpha(k)} \zeta^{\alpha} +
\bar\gamma_{k-1}f^{\alpha(k-1)\dot\alpha(k-1)} \zeta^{\dot\alpha}
\nonumber
\\
 && + \bar\beta'_{k} \Omega^{\alpha(k-1)\dot\alpha(k)\dot\beta}
\zeta_{\dot\beta} + \bar\gamma'_{k}
f^{\alpha(k-1)\beta\dot\alpha(k)}\zeta_{\beta}. \nonumber
\end{eqnarray}
where all coefficients are complex. As we will see below
coefficients in the supertransformations can be pure real or pure
imaginary. It depends on a parity of bosonic fields that is on how
bosonic fields transform under spatial reflections. The parity is
defined by operator $P$, acting on bosonic fields it gives
$$
Pf^{\alpha(k-1)\dot\alpha(k-1)}= \pm
f^{\alpha(k-1)\dot\alpha(k-1)},\quad
P\Omega^{\alpha(k)\dot\alpha(k-2)}=\pm
\Omega^{\alpha(k-2)\dot\alpha(k)}.
$$
The $+$, $-$ signs define parity-even and parity-odd bosonic fields
respectively. Considering fermionic fields
$\Phi^{\alpha(k),\dot\alpha(k-1)}$ and parameter of
supertransformations $\zeta^\alpha$ as parity-even, one can see that
in the case of parity-even(odd) bosonic fields coefficients
$\alpha_k,\alpha'_k$ are imaginary(real) and
$\beta_k,\beta'_k,\gamma_k,\gamma'_k$ are real(imaginary). As in the
case of the massive supermultiplets, partially massless ones have to
contain two bosonic fields with opposite parities since it arises from
the massive one in partially massless limit. Hence we have to consider
partially massless superblocks with parity-even bosonic field as
well as parity-odd one. So to unify these two cases we begin with
complex coefficients in supertransformations (\ref{PMST}).

In the gauge invariant formulation that we use it is easy to see
that not only spins but the depths of the superpartners must be
related. Indeed, let us consider partially massless bosonic field
$f_{[s]_t}$ of spin $s$ and depth $t=(s-l)$, which involves the
field variables $f^{\alpha(k-1)\dot\alpha(k-1)}$ with $l\leq k\leq
s$, i. e. it have maximal helicity $s$ and minimal one $l$. Then
there are only four possible superpartners, namely, partially
massless fermions with maximal helicities $s \pm 1/2$ and minimal
ones $l \pm 1/2$. Denoting partially massless fermionic field of
spin $s+1/2$ and depth $t=(s-l)$ as $\Phi_{[s+\frac12]_t}$, which
involves the field variables $\Phi^{\alpha(k)\dot\alpha(k-1)}$ with
$l\leq k\leq s$, four possible superpartners of bosonic field
$f_{[s]_t}$ can be represented by diagram
\begin{eqnarray}\label{4SB}
\xymatrix{\Phi_{[s+\frac12]_{t+1}} \ar@{-}[dr] &  &
\Phi_{[s+\frac12]_t}   \\
& f_{[s]_t} \ar@{-}[ur] &  \\
\Phi_{[s-\frac12]_{t}} \ar@{-}[ur] & & \Phi_{[s-\frac12]_{t-1}}
\ar@{-}[ul] }
\end{eqnarray} One can see from the diagram given above there are
just four superblocks that form partially massless supermultiplets.
Note that the ansatz for supertransformations (\ref{PMST}) is valid
for all types of the superblocks which differ only by the boundary
conditions.

Now let us consider a sum of the bosonic (\ref{BosonVar}) and
fermionic (\ref{FermVar}) variations under the supertransformations
(\ref{PMST}). Using the torsion zero conditions (\ref{PMOn-Shell}), it
can be written in the form $\delta{\cal L}+\delta{\cal L}'$, where
\begin{eqnarray}\label{PMVar}
\delta{\cal L} &=& \sum_{k=2}^s[- (-1)^\tau(k-1)\bar{\beta}_{k-1}
{\cal F}_{\alpha(k-1)\beta\dot\alpha(k-1)} e^\beta{}_{\dot\beta}
\Omega^{\alpha(k-2)\dot\alpha(k-1)\dot\beta} \zeta^{\alpha} \nonumber
\\
&& \qquad +(-1)^\sigma 4i\alpha_{k-1}
\Phi_{\alpha(k-2)\beta\gamma\dot\alpha(k-1)} e^\gamma{}_{\dot\gamma}
{\cal R}^{\alpha(k-2)\dot\alpha(k-1)\dot\gamma} \zeta^{\beta}
\nonumber
\\
&& \qquad -(-1)^\tau\bar{\gamma}_{k-1}
({\cal F}_{\alpha(k-1)\beta\dot\alpha(k-1)} e^\beta{}_{\dot\beta}
f^{\alpha(k-1)\dot\alpha(k-1)} \zeta^{\dot\beta} \nonumber
\\
&& \qquad + (k-1){\cal F}_{\alpha(k-1)\beta\dot\alpha(k-1)}
e^\beta{}_{\dot\beta}
f^{\alpha(k-1)\dot\alpha(k-2)\dot\beta}\zeta^{\dot\alpha})] + h.c.,
\\
\delta{\cal L}' &=& \sum_{k=2}^{s+1}[-(-1)^{\tau}\bar{\beta}'_{k-1}
{\cal F}_{\alpha(k-2)\gamma\dot\alpha(k-2)} e^\gamma{}_{\dot\gamma}
\Omega^{\alpha(k-2)\dot\alpha(k-2)\dot\gamma\dot\beta}
\zeta_{\dot\beta} \nonumber
\\
&& \qquad -(-1)^{\sigma}i4(k-1)\alpha'_{k-1}
\Phi_{\alpha(k-2)\beta\dot\alpha(k-2)} e^\beta{}_{\dot\beta}
{\cal R}^{\alpha(k-2)\dot\alpha(k-2)\dot\beta\dot\gamma}
\zeta_{\dot\gamma} \nonumber
\\
&& \qquad -(-1)^{\tau}\bar{\gamma}'_{k-1}
{\cal F}_{\alpha(k-2)\gamma\dot\alpha(k-2)} e^\gamma{}_{\dot\gamma}
f^{\alpha(k-2)\beta\dot\alpha(k-2)\dot\gamma} \zeta_\beta] +
h.c.\nonumber
\end{eqnarray}
Schematically, the structure of the variations has the form "curvature
$\times$ field". The fact, that both the Lagrangians and their
variations are defined only up to a total derivative, leads to a
number of non-trivial identities on such terms. The general form of
these identities were given in Appendix A of \cite{BKSZ18a} and they
are applicable to the case at hands, the only difference being in the
explicit expressions for the coefficients $a_k$, $b_k$, $c_k$ and
$d_k$. Using these identities one can express the parameters $\alpha$
and $\gamma$ in terms of $\beta$:
\begin{eqnarray}\label{Alf}
\alpha_k = (-1)^{\sigma+\tau}i \frac{k}{4}\bar\beta_k,\qquad
\alpha'_k = - (-1)^{\sigma+\tau}\frac{i}{4k}
\bar\beta'_{k-1},
\end{eqnarray}
\begin{eqnarray}\label{Gam}
\gamma_k &=& 2d_{k+1} \bar\beta_k,\qquad
\gamma'_k = 2d_k \bar\beta'_k.
\end{eqnarray}
Also we obtain recurrence relations on the parameters $\beta_k$ and
$\beta'_k$:
\begin{eqnarray}\label{RecurEq}
2(k+1)\beta_{k-1} c_{k+1} = k\beta_{k} a_{k+1}, \qquad \beta'_{k-1}
a_{k+1} = 2\beta'_kc_k.
\end{eqnarray}
Last but not least, we obtain four independent equations which relate
$\beta$ and $\beta'$ as well as the bosonic and fermionic depth
parameters:
\begin{eqnarray}
0 &=& \frac{\beta'_{k-1}c_k}{(k-1)} - \frac{\beta'_ka_{k+1}}{2(k+1)}
+ \lambda \beta_{k-1} - \gamma_{k-1}, \label{Relat1}
\\
0 &=& (k-1)\beta_{k-1}c_{k} - \frac{(k-2)}{2} \beta_{k-2}a_{k} +
\lambda \beta'_{k-1} - \gamma'_{k-1}, \label{Relat2}
\\
0 &=& \frac{(k-1)}{2k} \bar\beta_{k-1}b_{k} - 2kd_{k}\gamma_{k-1}
+\lambda \bar\gamma_{k-1} -
\frac{\bar\gamma'_{k}a_{k+1}}{2k(k+1)},\label{Relat3}
\\
0 &=& \frac{\bar\beta'_{k-1}b_{k}}{2} - 2(k-1)d_{k-1}\gamma'_{k-1}
-\lambda\bar\gamma'_{k-1} - \bar\gamma_{k-1}c_k. \label{Relat4}
\end{eqnarray}
In the next two subsections we find the solutions of these equations
for the two possible the partially massless superblocks.

\subsection{Solution for the superblock $(s+1/2,s)$}\label{PMSB1}

Let us consider a superblock containing a partially massless boson
with
spin $s$ and depth $t = (s-l-1)$ and a partially massless fermion with
spin $s+1/2$ and depth $\tilde{t} = (s-\tilde{l}-1$. The explicit
expressions for the bosonic coefficients have the form:
\begin{eqnarray*}
b_k &=& \frac{2s(s+1)l(l+1)}{k(k-1)(k+1)}\lambda^2, \nonumber
\\
a_k{}^2 &=& \frac{4(s-k+1)(s+k)(k-l-1)(k+l)}{(k-2)(k-1)}\lambda^2,
\end{eqnarray*}
while the fermionic ones look like:
\begin{eqnarray*}
d_k &=& \pm\frac{(s+1)(\tilde{l}+1)}{2k(k+1)}\lambda, \nonumber
\\
c_k{}^2 &=&
\frac{(s-k+1)(s+k+1)(k-\tilde{l}-1)(k+\tilde{l}+1)}{k^2}\lambda^2.
\end{eqnarray*}
Recall that the parameters $\alpha_k$ and $\gamma_k$ are determined in
terms of $\beta$ by (\ref{Alf}) and (\ref{Gam}). Now let us consider
equation (\ref{Relat3}) at $k=s$. This gives
$$
[l(l+1)-(\tilde{l}+1)^2]\bar{\beta}_{s-1} =
\mp(\tilde{l}+1)\beta_{s-1},
$$
where the sign corresponds to that of $d_k$, and provides us with the
relation on the bosonic and fermionic depths:
\begin{eqnarray*}
\tilde{l} &=& l, \qquad \bar\beta_{s-1} = \pm\beta_{s-1},
\\
\tilde{l} &=& l-1, \quad \bar\beta_{s-1} = \mp \beta_{s-1}.
\end{eqnarray*}
Remind that a real (imaginary) values of $\beta_{s-1}$ corresponds
to a parity-even (parity-odd) bosonic field. Now we proceed with the
solution of all remaining equations and obtain, for $\tilde{l}=l$
\begin{eqnarray}\label{solBet1}
\beta_k = \sqrt{\frac{(s+k+2)(k+l+2)}{k}}\beta, \quad
\beta'_k = - \sqrt{{k(s-k)(k-l)}}\beta,
\end{eqnarray}
and for $\tilde{l}=l-1$
\begin{eqnarray}\label{solBet2}
\beta_k = \sqrt{\frac{(s+k+2)(k-l+1)}{k}}\beta, \quad
\beta'_k = - \sqrt{k(s-k)(k+l+1)}\beta,
\end{eqnarray}
where $\beta=\rho$ ($\beta=i\rho$) takes pure real (imaginary)
value. These four solutions corresponds to upper line in diagram
(\ref{4SB}) and schematically can be presented as
\begin{eqnarray}\label{SolutSB}
\xymatrix { {[s+\frac12]}_{t}^{\tau,\pm} \ar@{-}[d]_-{\rho} &
 \\
{[s]}_{t}^{\sigma,\pm} & }
 \qquad
\xymatrix {  {[s+\frac12]}_{t+1}^{\tau,\mp}\ar@{-}[d]_-{\rho} &
 \\
{[s]}_{t}^{\sigma,\pm} & }
\end{eqnarray}
with following clarifying features. $\pm$ signs for the bosons
correspond to their parity, while for the fermions to the sign of
$d_k$. The $\sigma,\tau$ parameterize norm of kinetic terms in
Lagrangian for bosons (\ref{PMBosonLag}) and fermions
(\ref{FermLag}) respectively, they are still unfixed at this stage.
The real parameter $\rho$ corresponds to one free parameter in
supertransformations. Each superblocks must have its own parameter 
$\rho$.

\subsection{Solution for the superblock $(s,s-1/2)$}\label{PMSB2}

Now let us turn to the second superblock which contain a partially
massless boson with spin $s$ and depth $t=(s-l-1)$ and a partially
massless fermion with spin $s-1/2$ and depth
$\tilde{t} = (s-\tilde{l}-2)$. Thus for the bosonic field we still
have the same expressions for the coefficients $a_k$ and $b_k$ as in
the previous subsection, while for the fermion we obtain
\begin{eqnarray*}
d_k &=& \pm\frac{s(\tilde{l}+1)}{2k(k+1)}\lambda,
\\
c_k{}^2 &=&
\frac{(s-k)(s+k)(k-\tilde{l}-1)(k+\tilde{l}+1)}{k^2}\lambda^2.
\end{eqnarray*}
First of all, let us consider equation (\ref{Relat2}) at $k=s$. This
gives us:
$$
[l(l+1)-(\tilde{l}+1)^2]\beta'_{s-1} =
\pm(\tilde{l}+1)\bar{\beta}'_{s-1}.
$$
where the sign corresponds to that of $d_k$. Thus in this case we
again have four possible solutions:
\begin{eqnarray*}
\tilde{l} &=& l, \qquad \bar\beta_{s-1} = \mp\beta_{s-1},
\\
\tilde{l} &=& l-1, \quad \bar\beta_{s-1} = \pm \beta_{s-1}.
\end{eqnarray*}
Then the solution of the remaining equations gives, for $\tilde{l} =
l$
\begin{eqnarray}\label{solBet3}
\beta_k = \sqrt{\frac{(s-k-1)(k+l+2)}{k}}\beta, \quad
\beta'_k = \sqrt{k(s+k+1)(k-l)}\beta,
\end{eqnarray}
and for $\tilde{l} = l-1$
\begin{eqnarray}\label{solBet4}
\beta_k = \sqrt{\frac{(s-k-1)(k-l+1)}{k}}\beta, \quad
\beta'_k = \sqrt{k(s+k+1)(k+l+1)}\beta,
\end{eqnarray}
where again $\beta=\rho$ ($\beta=i\rho$) takes pure real (imaginary)
value. These four solutions corresponds to lower line in diagram
(\ref{4SB}) and schematically can be presented as
\begin{eqnarray}\label{SolutSB2}
\xymatrix {  {[s]}_{t}^{\sigma,\pm} \ar@{-}[d]_-{\rho} &
 \\
{[s-\frac12]}_{t}^{\tau,\pm} & }
 \qquad
\xymatrix { {[s]}_{t}^{\sigma,\pm} \ar@{-}[d]_-{\rho} &
 \\
{[s-\frac12]}_{t-1}^{\tau,\mp} & }
\end{eqnarray}
Here all additional notations are the same as in previous case
(\ref{SolutSB}).

\section{Partially massless supermultiplets}\label{Section4}

As we have already mentioned, each partially massless
supermultiplets contains two bosonic and two fermionic fields. As in
the massive case, the two bosons must have opposite parities, and it
turns out to be important that the two fermions have opposite signs
of their mass terms. Moreover, the depths of the partial
masslessness for each field must be properly adjusted.
Schematically, the two possible supermultiplets (\ref{PMSM}) look
like:
$$
\xymatrix {& {[s+\frac12]_{t}^{\tau_1,+}} \ar@{-}[dl]_-{\rho_1}
\ar@{-}[dr]^-{\rho_3} &\\
 {[s]_{t}^{\sigma_1,+}} \ar@{-}[dr]_-{\rho_2} & Y=s &
{[{s}]_{t-1}^{\sigma_2,-}}\\
 & {[s-\frac12]_{t-1}^{\tau_2,-}} \ar@{-}[ur]_-{\rho_4} & }
\qquad \xymatrix {& {[s-\frac12]_{t}^{\tau_1,+}} \ar@{-}[dl]_-{\rho_1}
\ar@{-}[dr]^-{\rho_3} &\\
 {[s]_{t}^{\sigma_1,+}} \ar@{-}[dr]_-{\rho_2} & Y=s-\frac12 &
{[{s-1}]_{t-1}^{\sigma_2,-}}\\
 & {[s-\frac12]_{t-1}^{\tau_2,-}} \ar@{-}[ur]_-{\rho_4} & }
$$
with the same additional notations that was used in the construction
of superblocks (\ref{SolutSB}), (\ref{SolutSB2}). Since these
supermultiplets are constructed as combination of superblocks we
should take independent parameters $\sigma_i,\tau_i$ which normalize
kinetic terms for bosons and fermions respectively and independent
$\rho_i$ which parameterize supertransformations for given
superblock.

Let us use the notations $(f_+,\Omega_+)$ and $(f_-,\Omega_-)$ for the
parity-even/parity-odd bosons and $\Phi_+,\Phi_-$ for fermions
according to their sign of $d_k$. In these notations the
supertransformations for the whole supermultiplets are the combination
of four separate superblocks corresponding to the parameters
$\rho_{1,2,3,4}$ shown above. Namely, we take for the bosons:
\begin{eqnarray*}
\delta f_+^{\alpha(k-1)\dot\alpha(k-1)} &=&
\alpha_{k-1}|_{\rho_1}\Phi_+^{\alpha(k-1)\beta\dot\alpha(k-1)}
\zeta_\beta + \alpha'_{k-1}|_{\rho_1}
\Phi_+^{\alpha(k-1)\dot\alpha(k-2)}\zeta^{\dot\alpha}
\\
 && + \alpha_{k-1}|_{\rho_2}
\Phi_-^{\alpha(k-1)\beta\dot\alpha(k-1)}\zeta_\beta
 + \alpha'_{k-1}|_{\rho_2}
\Phi_-^{\alpha(k-1)\dot\alpha(k-2)}\zeta^{\dot\alpha} + h.c.
\end{eqnarray*}
and similarly for $\delta f_-^{\alpha(k-1)\dot\alpha(k-1)}$ with the
replacement $\rho_1 \to \rho_3$ and $\rho_2 \to \rho_4$, while for the
fermions we use
\begin{eqnarray*}
\delta \Phi_+^{\alpha(k)\dot\alpha(k-1)} &=& \beta_{k-1}|_{\rho_1}
\Omega_+^{\alpha(k)\dot\alpha(k-2)} \zeta^{\dot\alpha} +
\gamma_{k-1}|_{\rho_1} f_+^{\alpha(k-1)\dot\alpha(k-1)}
\zeta^{\alpha}
\\
 && + \beta'_{k}|_{\rho_1}
\Omega_+^{\alpha(k)\beta\dot\alpha(k-1)}\zeta_\beta +
\gamma'_{k}|_{\rho_1} f_+^{\alpha(k)\dot\alpha(k-1)\dot\beta}
\zeta_{\dot\beta}
\\
&& + \beta_{k-1}|_{\rho_3} \Omega_-^{\alpha(k)\dot\alpha(k-2)}
\zeta^{\dot\alpha} + \gamma_{k-1}|_{\rho_3}
f_-^{\alpha(k-1)\dot\alpha(k-1)} \zeta^{\alpha}
\\
 && + \beta'_{k}|_{\rho_3}
\Omega_-^{\alpha(k)\beta\dot\alpha(k-1)}\zeta_\beta +
\gamma'_{k}|_{\rho_3} f_-^{\alpha(k)\dot\alpha(k-1)\dot\beta}
\zeta_{\dot\beta}
\end{eqnarray*}
and similarly for $\delta \Phi_-^{\alpha(k)\dot\alpha(k-1)}$  with the
replacement $\rho_1 \to \rho_2$ and $\rho_3 \to \rho_4$.

So to construct a complete partially massless supermultiplet we have
to adjust these four parameters $\rho_{1,2,3,4}$ so that the algebra
of supertransformations be closed. It means that the commutator of the
two supertransformations must produce a combination of translations
and Lorentz transformations:
\begin{equation}\label{SA}
\{Q_\alpha,Q_{\dot\alpha}\} \sim P_{\alpha\dot\alpha}, \quad
\{Q_\alpha,Q_{\alpha}\} \sim \lambda M_{\alpha\alpha}, \quad
\{Q_{\dot\alpha},Q_{\dot\alpha}\} \sim \lambda
M_{\dot\alpha\dot\alpha}.
\end{equation}
The structure of the mass-shell condition (\ref{PMOn-Shell}) shows
that, for example, the commutator on the bosonic field
$f_+{}^{\alpha(k-1)\dot\alpha(k-1)}$ must only contain
$\Omega_+{}^{\alpha(k)\dot\alpha(k-2)}$,
$\Omega_+{}^{\alpha(k-2)\dot\alpha(k)}$,
$f_+^{\alpha(k)\dot\alpha(k)}$, $f_+{}^{\alpha(k-1)\dot\alpha(k-1)}$
and $f_+{}^{\alpha(k-2)\dot\alpha(k-2)}$. This requirement leads to
the
number of relations on the parameters:
$$
\alpha_{k-1}|_{\rho_1} \beta'_{k}|_{\rho_1} + \alpha_{k-1}|_{\rho_2}
\beta'_{k}|_{\rho_2} = 0, \qquad \alpha'_{k-1}|_{\rho_1}
\beta_{k-2}|_{\rho_1} + \alpha'_{k-1}|_{\rho_2}
\beta_{k-2}|_{\rho_2} = 0,
$$
$$
\alpha_{k-1}|_{\rho_1} \beta'_{k}|_{\rho_3} +
\alpha_{k-1}|_{\rho_2} \beta'_{k}|_{\rho_4} = 0, \qquad
\alpha'_{k-1}|_{\rho_1} \beta_{k-2}|_{\rho_3} +
\alpha'_{k-1}|_{\rho_2} \beta_{k-2}|_{\rho_4} = 0,
$$
$$
\alpha_{k-1}|_{\rho_1} \beta_{k-1}|_{\rho_3}
+ \alpha'_{k-1}|_{\rho_1} \beta'_{k-1}|_{\rho_3} +
\alpha_{k-1}|_{\rho_2} \beta_{k-1}|_{\rho_4}
+ \alpha'_{k-1}|_{\rho_2} \beta'_{k-1}|_{\rho_4} = 0,
$$
$$
\alpha_{k-1}|_{\rho_1} \gamma_{k-1}|_{\rho_3}
- \bar\alpha'_{k-1}|_{\rho_1} \bar\gamma'_{k-1}|_{\rho_3} +
\alpha_{k-1}|_{\rho_2} \gamma_{k-1}|_{\rho_4}
- \bar\alpha'_{k-1}|_{\rho_2} \bar\gamma'_{k-1}|_{\rho_4} = 0,
$$
$$
\alpha_{k-1}|_{\rho_1} \gamma'_{k}|_{\rho_3}
- \bar\alpha_{k-1}|_{\rho_1} \bar\gamma'_{k}|_{\rho_3} +
\alpha_{k-1}|_{\rho_2} \gamma'_{k}|_{\rho_4}
- \bar\alpha_{k-1}|_{\rho_2} \bar\gamma'_{k}|_{\rho_4} = 0,
$$
$$
\alpha'_{k-1}|_{\rho_1} \gamma_{k-2}|_{\rho_3}
- \bar\alpha'_{k-1}|_{\rho_1} \bar\gamma_{k-2}|_{\rho_3} +
\alpha'_{k-1}|_{\rho_2} \gamma_{k-2}|_{\rho_4}
- \bar\alpha'_{k-1}|_{\rho_2} \bar\gamma_{k-2}|_{\rho_4} = 0.
$$
If these relations are fulfilled, the general form of the
commutator looks like:
\begin{eqnarray}\label{GCom}
\ [\delta_1, \delta_2 ] f_+^{\alpha(k-1)\dot\alpha(k-1)} &=&
 (\alpha_{k-1}|_{\rho_1} \beta_{k-1}|_{\rho_1}
+ \alpha'_{k-1}|_{\rho_1} \beta'_{k-1}|_{\rho_1} +
\alpha_{k-1}|_{\rho_2} \beta_{k-1}|_{\rho_2} +
\alpha'_{k-1}|_{\rho_2} \beta'_{k-1}|_{\rho_2})\nonumber
\\
 && \cdot [ \xi^\alpha{}_{\dot\beta}
\Omega_+^{\alpha(k-2)\dot\alpha(k-1)\dot\beta} +
\xi_\beta{}^{\dot\alpha} \Omega_+^{\alpha(k-1)\beta\dot\alpha(k-2)}
]\nonumber
\\
 && + (\alpha_{k-1}|_{\rho_1} \gamma'_k|_{\rho_1}
+ \alpha_{k-1}|_{\rho_2} \gamma'_k|_{\rho_2})
f_+^{\alpha(k-1)\beta\dot\alpha(k-1)\dot\beta}
\xi_{\beta\dot\beta}
\\
 && + (\alpha'_{k-1}|_{\rho_1} \gamma_{k-2}|_{\rho_1}
+ \alpha'_{k-1}|_{\rho_2} \gamma_{k-2}|_{\rho_2})
f_+^{\alpha(k-2)\dot\alpha(k-2)} \xi^{\alpha\dot\alpha}\nn
\\
 && + (\alpha_{k-1}|_{\rho_1} \gamma_{k-1}|_{\rho_1}
 + \alpha'_{k-1}|_{\rho_1} \gamma'_{k-1}|_{\rho_1}
 + \alpha_{k-1}|_{\rho_2} \gamma_{k-1}|_{\rho_2}
 + \alpha'_{k-1}|_{\rho_2} \gamma'_{k-1}|_{\rho_2})\nn
\\
 && \cdot [ f_+^{\alpha(k-2)\beta\dot\alpha(k-1)}
\eta^\alpha{}_\beta + f_+^{\alpha(k-1)\dot\alpha(k-2)\dot\beta}
\eta^{\dot\alpha}{}_{\dot\beta}], \nn
\end{eqnarray}
where
$$
\xi^{\alpha\dot\alpha} = \zeta_1^\alpha \zeta_2^{\dot\alpha} -
\zeta_2^\alpha \zeta_1^{\dot\alpha}, \qquad
\eta^{\alpha(2)} = \zeta_1^\alpha \zeta_2^\alpha - \zeta_2^\alpha
\zeta_1^\alpha, \qquad \eta^{\dot\alpha(2)} = \zeta_1^{\dot\alpha}
\zeta_2^{\dot\alpha} - \zeta_2^{\dot\alpha} \zeta_1^{\dot\alpha}.
$$
For the bosonic field $f_-{}^{\alpha(k-1)\dot\alpha(k-1)}$ the
commutator has the same form with the replacements
$\rho_1\rightarrow\rho_3$ and $\rho_2\rightarrow\rho_4$. Let us
stress that all these bosonic components belong to the same
supermultiplet, i.e. to the same irreducible representations. Thus
all the expressions in round brackets in (\ref{GCom}) must be
$k$-independent. This gives additional restrictions on the
parameters and also serves as quite a non-trivial test for our
calculations.

\subsection{Supermultiplets with half-integer superspin}

The partially massless supermultiplet with the half-integer
superspin $Y=(s-1/2)$ has the following structure:
$$
\xymatrix {& {[s-\frac12]_{t}^{\tau_1,+}} \ar@{-}[dl]_-{\rho_1}
\ar@{-}[dr]^-{\rho_3} &\\
 {[s]_{t}^{\sigma_1,+}} \ar@{-}[dr]_-{\rho_2} & Y=s-\frac12 &
{[{s-1}]_{t-1}^{\sigma_2,-}}\\
 & {[s-\frac12]_{t-1}^{\tau_2,-}} \ar@{-}[ur]_-{\rho_4} & }
$$
We have only a handful of free parameters in our disposal and a large
number of equations to fulfill, however, the closure of the
superalgebra is achieved at:
\begin{equation}
\sigma_2 = \sigma_1, \quad \tau_2 = \tau_1+1, \quad \rho_1{}^2 =
\rho_2{}^2 = \rho_3{}^2 = \rho_4{}^2, \qquad \rho_1\rho_3 =
\rho_2\rho_4. \label{ConFer}
\end{equation}
Note that such relations between $\sigma,\tau$ parameters mean that
two bosons must enter with opposite norms of kinetic terms and the
same is for two fermions\footnote{ We recall that work in metric
signature $(+,-,-,-)$ which give overall $(-1)^s$ factor in norm of
kinetic terms for given bosonic field with spin $s$. So the same
factor for bosonic field with spin $(s-1)$ means opposite norm of
kinetic terms $(-1)^s=-(-1)^{s-1}$. This explains unusual relation
between $\sigma_1,\sigma_2$ in (\ref{ConFer})}. This is in agreement
with \cite{GHR18} where the same result was obtained for the case of
the supermultiplet with partially massless spin-$2$ field.

The final form for the commutators of the supertransformations on
parity-even spin-$s$ $f_+$ and parity-odd spin-$(s-1)$ $f_-$ fields
appears to be the same:
\begin{eqnarray*}
\frac{1}{i\rho_0{}^2}{[\delta_1,\delta_2]}
f_\pm^{\alpha(k-1)\dot\alpha(k-1)} &=&
\Omega_\pm^{\alpha(k-1)\beta\dot\alpha(k-2)} \xi_\beta{}^{\dot\alpha}
+ \Omega_\pm^{\alpha(k-2)\dot\alpha(k-1)\dot\beta}
\xi^\alpha{}_{\dot\beta}
\\
 && - \frac{(k-1)a_{k+1}}{2(k+1)}
f_\pm^{\alpha(k-1)\beta\dot\alpha(k-1)\dot\beta} \xi_{\beta\dot\beta}
 + \frac{a_{k}}{2k(k-1)} f_\pm^{\alpha(k-2)\dot\alpha(k-2)}
\xi^{\alpha\dot\alpha}
\\
 && +\lambda[
f_\pm^{\alpha(k-2)\beta\dot\alpha(k-1)}(\zeta_1^{\alpha}
\eta^\alpha{}_\beta + f_\pm^{\alpha(k-1)\dot\alpha(k-2)\dot\beta}
\eta^{\dot\alpha}{}_{\dot\beta}],
\end{eqnarray*}
where $a_k$ is given by (\ref{boson_date}) for spin $s$ and spin
$(s-1)$ respectively and
$$
\rho_0{}^2 = - (-1)^{\sigma_1+\tau_1}\frac{s(2l+1)}{2}\rho_1{}^2.
$$

\subsection{Supermultiplets with integer superspin}

Now let us turn to the partially massless supermultiplet with integer
superspin $Y=s$:
$$
\xymatrix {& {[s+\frac12]_{t}^{\tau_1,+}} \ar@{-}[dl]_-{\rho_1}
\ar@{-}[dr]^-{\rho_3} &\\
 {[s]_{t}^{\sigma_1,+}} \ar@{-}[dr]_-{\rho_2} & Y=s &
{[{s}]_{t-1}^{\sigma_2,-}}\\
 & {[s-\frac12]_{t-1}^{\tau_2,-}} \ar@{-}[ur]_-{\rho_4} & }
$$
All the relations for the closure of the superalgebra are fulfilled
provided:
\begin{equation}
\sigma_2 = \sigma_1+1, \quad \tau_2 = \tau_1, \quad \rho_1{}^2 =
\rho_2{}^2 = \rho_3{}^2 = \rho_4{}^2, \quad \rho_1\rho_3 =
\rho_2\rho_4. \label{ConBos}
\end{equation}
Again we see that such relations between $\sigma,\tau$ parameters
mean that two bosons must enter with opposite norm of kinetic terms
and the same is for two fermions\footnote{ We again recall that work
in metric signature $(+,-,-,-)$ which give overall $(-1)^s$ factor
in norm of kinetic terms for given fermionic field with spin
$s+1/2$. So the same factor for fermionic field with spin $s-1/2$
means opposite norm of kinetic terms $(-1)^s=-(-1)^{s-1}$. This
explains unusual relation between $\tau_1,\tau_2$ in
(\ref{ConBos})}.

The final form of the commutators of the supertransformations on
parity-even and parity-odd bosonic spin-$s$ fields have the same form
as in the previous case:
\begin{eqnarray*}
\frac{1}{i\rho_0{}^2}{[\delta_1,\delta_2]}
f_\pm^{\alpha(k-1)\dot\alpha(k-1)} &=&
\Omega_\pm^{\alpha(k-1)\beta\dot\alpha(k-2)} \xi_\beta{}^{\dot\alpha}
 + \Omega_\pm^{\alpha(k-2)\dot\alpha(k-1)\dot\beta}
\xi^\alpha{}_{\dot\beta}
\\
 && - \frac{(k-1)a_{k+1}}{2(k+1)}
f_\pm^{\alpha(k-1)\beta\dot\alpha(k-1)\dot\beta}
\xi_{\beta\dot\beta}  + \frac{a_{k}}{2k(k-1)}
f_\pm^{\alpha(k-2)\dot\alpha(k-2)} \xi^{\alpha\dot\alpha}
\\
 && +\lambda[ f_\pm^{\alpha(k-2)\beta\dot\alpha(k-1)}
\eta^\alpha{}_\beta + f_\pm^{\alpha(k-1)\dot\alpha(k-2)\dot\beta}
\eta^{\dot\alpha}{}_{\dot\beta}],
\end{eqnarray*}
where $a_k$ is given by (\ref{boson_date}) for spin $s$ and
$$
\rho_0{}^2 = (-1)^{\sigma_1+\tau_1} \frac{(2s+1)(l+1)}{2}\rho_1{}^2.
$$

\section{Summary}

In this paper we have presented the component Lagrangian description
of partially massless higher spin on-shell arbitrary $N=1$
supermultiplets in four-dimensional $AdS_4$ space corresponding the
classification given in \cite{GHR18}\footnote{In higher dimensions
there exists much more rich spectrum of the partially massless fields
including mixed symmetry ones. So, in principle, there may exists a
whole zoo of the corresponding supermultiplets. However, as far as we
know, such supersymmetric representations are not studied at
present.}. The constructed supermultiplets are
non-unitary and contain partially massless fields with appropriately
chosen spins and depths. As in a massive case \cite{BKSZ18a} we show
that $N=1$ partially massless supermultiplets can be constructed as
a combination of four partially massless superblocks containing one
partially massless boson and one partially massless fermion. As a
result we have derived both the supertransformations for the
components of the supermultiplet and the corresponding invariant
Lagrangian. Also we show that a closure of superalgebra requires
that two bosons and two fermions must enter supermultiplets with
opposite norm of kinetic terms. All our results are in  agreement
with the results of \cite{GHR18} and extend them. The constructed
Lagrangian formulation describes a dynamics of arbitrary superspin
partially massless supermultiplets in $AdS_4$ space.

\section*{Acknowledgments}
I.L.B and T.V.S are grateful to the RFBR grant, project No.
18-02-00153-a for partial support. Their research was also supported
in parts by Russian Ministry of Science and High Education, project
No. 3.1386.2017. T.V.S acknowledges partial support from the
President of Russia grant for young scientists No. MK-1649.2019.2.

\appendix

\section{Notations and conventions}
We work in the frame-like multispinor formalism. It means that all
objects are differential p-forms ($p=0,1,2,3,4$ in four dimensions)
with multispinors as their local indices, i.e.
$$
\Omega^{\alpha(m)\dot\alpha(n)}=dx^{\mu_1}\wedge...\wedge
dx^{\mu_p}\Omega_{\mu_1...\mu_p}{}^{\alpha(m)\dot\alpha(n)}.
$$
Here $dx^\mu$ are required to anti-commute $dx^{\mu}\wedge
dx^{\nu}=-dx^{\nu}\wedge dx^{\mu}$ with respect to exterior product
$\wedge$. World indices $\mu,\nu$ are omitted everywhere; all
expressions are completely antisymmetric on them. We use the
condensed notations for local multispinor indices
$\alpha,\dot\alpha$. Namely, all objects are totally symmetric on
upper/low undotted/dotted indices $\alpha_1\alpha_2\cdots\alpha_k$,
we denote them by the same letter with the number of indices in
parentheses. For example:
$$
\Omega^{(\alpha_1\alpha_2 \dots \alpha_{m})(\dot\alpha_1\dot\alpha_2
\dots \dot\alpha_{n})}=\Omega^{\alpha(m)\dot\alpha(n)}.
$$
We also always assume if in expression spinor indices denoted by the
same letters and placed on the same level are symmetrized, e.g.
$$
\Omega^{\alpha(m)} \zeta^{\alpha} = \Omega^{(\alpha_1\dots
\alpha_{m}} \zeta^{\dot\alpha_{m+1})}= \Omega^{\alpha_1\dots
\alpha_{m}} \zeta^{\dot\alpha_{m+1}}+ \mbox{permutations ($m$
terms)}.
$$
The spinor indices are raised and lowered with the antisymmetric
tensors $ \epsilon_{\alpha\beta}$
($\epsilon_{\dot\alpha\dot\beta}$):
\begin{equation}
 \epsilon_{\alpha\beta} \xi^\beta = - \xi_\alpha, \qquad
 \epsilon^{\alpha\beta} \xi_\beta = \xi^\alpha,
\end{equation}
the same is true for dotted indices. Hence, all the symmetric
multispinors are automatically traceless. Under the Hermitian
conjugation, dotted and undotted indices are transformed one into
another. For example:
$$
\left( \Omega^{\alpha(m)\dot\alpha(n)}\right)^\dagger =
\Omega^{\alpha(n)\dot\alpha(m)}.
$$

The $AdS_4$ space is described by the background Lorentz connections
$\omega^{\alpha(2)}$, $\omega^{\dot\alpha(2)}$, which enter
implicitly through the Lorentz covariant derivative $D$, and the
background frame $e^{\alpha\dot\alpha}$. We also use the basis
elements for the two-, thee- and four-forms
\begin{equation}
 e^a \sim e^{\alpha\dot\alpha}, \qquad
 E^{ab} \sim E^{\alpha(2)}, E^{\dot\alpha(2)}, \qquad
 E^{abc} \sim E^{\alpha\dot\alpha}, \qquad
 E^{abcd} \sim E,
\end{equation}
defined as follows:
\begin{eqnarray}
e^{\alpha\dot\alpha} \wedge e^{\beta\dot\beta} &=&
\varepsilon^{\alpha\beta} E^{\dot\alpha\dot\beta} +
\varepsilon^{\dot\alpha\dot\beta} E^{\alpha\beta}, \nonumber
\\
E^{\alpha(2)} \wedge e^{\beta\dot\alpha} &=&
\varepsilon^{\alpha\beta}
E^{\alpha\dot\alpha}, \\
E^{\alpha\dot\alpha} \wedge e^{\beta\dot\beta} &=&
\varepsilon^{\alpha\beta} \varepsilon^{\dot\alpha\dot\beta} E.
\nonumber
\end{eqnarray}
The hermitian conjugation rules for the basis forms are:
\begin{equation}
\left( e^{\alpha\dot\alpha}\right)^\dagger = e^{\alpha\dot\alpha},
\qquad \left( E^{\alpha(2)}\right)^\dagger = E^{\dot\alpha(2)},
\qquad \left( E^{\alpha\dot\alpha}\right)^\dagger = -
E^{\alpha\dot\alpha}, \qquad \left(E\right)^\dagger = - E.
\end{equation}
The Lorentz covariant derivative is normalized so that
\begin{equation}
D\wedge D \Omega^{\alpha(m)\dot\alpha(n)} = -
2\lambda^2(E^\alpha{}_\beta\wedge\Omega^{\alpha(m-1)\beta\dot\alpha(n)}+E^{\dot\alpha}{}_{\dot\beta}
\wedge\Omega^{\alpha(m)\dot\alpha(n-1)\dot\beta}).
\end{equation}
The parameter $\lambda^2$ is proportional to the curvature of the
space-time. The $AdS$ space has $\lambda^2>0$, while the $d$S space
has $\lambda^2<0$. The case of $\lambda^2=0$ corresponds to the flat
Minkowski space.

In the main text all the wedge product signs $\wedge$ are omitted.


\begin{thebibliography}{10}

\bibitem{GHR18}
S. Garcia-Saenz, K. Hinterbichler, R.~A. Rosen, {\it Supersymmetric
Partially Massless Fields and Non-Unitary Superconformal
Representations,} JHEP {\bf 1811} (2018) 166 [arXiv:1810.01881].

\bibitem{DW01}
S.~Deser, A.~Waldron {\it "Gauge Invariance and Phases of Massive
Higher Spins in (A)dS",} Phys. Rev. Lett. {\bf 87} (2001) 031601,
hep-th/0102166.

\bibitem{DW01a}
S.~Deser, A.~Waldron {\it "Partial Masslessness of Higher Spins in
(A)dS",} Nucl. Phys. {\bf B607} (2001) 577, hep-th/0103198.

\bibitem{Zin01}
Yu.~M. Zinoviev
{\it "On Massive High Spin Particles in (A)dS",}
arXiv:hep-th/0108192.

\bibitem{Met06}
R.~R. Metsaev {\it "Gauge invariant formulation of massive totally
symmetric fermionic fields in (A)dS space",} Phys. Lett. {\bf B643}
(2006) 205-212, hep-th/0609029.

\bibitem{SV06}
E.~D.~Skvortsov, M.~A.~Vasiliev
{\it "Geometric formulation for partially massless fields",}
Nucl. Phys. {\bf B756} (2006) 117, arXiv:hep-th/0601095.

\bibitem{Zin08b}
Yu.~M. Zinoviev {\it "Frame-like gauge invariant formulation for
massive high spin particles",} Nucl. Phys. {\bf B808} (2009) 185,
[arXiv:0808.1778].

\bibitem{PV10}
D.~S. Ponomarev, M.~A. Vasiliev {\it "Frame-Like Action and Unfolded
Formulation for Massive Higher-Spin Fields",} Nucl. Phys. {\bf B839}
(2010) 466, arXiv:1001.0062.


\bibitem{BKSZ18a}
I.~L. Buchbinder, M.~V. Khabarov, T.~V. Snegirev, Yu.~M. Zinoviev,
{\it  Lagrangian formulation of the massive higher spin $N=1$
supermultiplets in $AdS_4$ space,} arXiv:1901.09637.

\bibitem{BK98}
I.~L.~Buchbinder, S.~M.~Kuzenko, {\it Ideas and Methods of
Supersymmetry and Supergravity,} IOP Publishing, Bristol and
Philadelphia, 1998.

\bibitem{BS02}
T. Biswas, W. Siegel, {\it Radial Dimensional Reduction: (Anti) de
Sitter Theories from Flat,} JHEP {\bf 0207} (2002) 005
[arXiv:hep-th/0203115].

\bibitem{HW05}
K. Hallowell, A. Waldron, {\it Constant Curvature Algebras and
Higher Spin Action Generating Functions,} Nucl. Phys. {\bf B724}
(2005) 453, [arXiv:hep-th/0505255].


\end{thebibliography}
\end{document}